\documentclass[]{IEEEphot}

\usepackage{amsmath,amssymb,amsfonts}
\usepackage{graphicx}

\jvol{xx}
\jnum{xx}
\jmonth{April}
\pubyear{2014}

\begin{document}

\title{Analysis of Cascaded Brillouin Scattering in Optical Fibres}
\author{Neil G.~R.~Broderick and Stephane X.~Coen }
\affil{Dodd-Walls Centre for Photonic and Quantum Technologies, Department of Physics, University of Auckland, Auckland NZ}

\maketitle

\markboth{IEEE Photonics Journal}{Cascaded Brillouin Scattering}

\begin{abstract}
 Cascaded Brillouin scattering is increasingly becoming of interest in many areas of photonics (e.g. see \cite{Liu:2012ek} and \cite{Buttner:2014cv})
 and has been studied experimentally by many groups. In gain assisted experiments for example up to nearly 800 distinct lines
have been observed\cite{BumkiMin:dt}  while in passive cavities, mode locking of a few Brillouin lines has recently been demonstrated\cite{Loranger:2012jn}. 
We show
that although each pair of lines interact through the creation of an acoustic phonon the resulting acoustic field nevertheless can
be written as the sum of only two waves (a forward and backward propagating wave). This new analysis results in a set of strongly
coupled amplitude equations for the fields that are derived here for the first time in the context of optical fibres. More-over we 
show that previous authors have mis-understood the nature of the acoustic field resulting in incorrect sets of coupled equations.
\end{abstract}


Stimulated Brillouin scattering (SBS) in transparent materials has long been studied due to its ubiquity 
and low threshold. In optical fibres for example the threshold can be as low as a few mW making it 
the dominant nonlinear process when dealing with the propagation of narrow linewidth sources\cite{agrawal}. In Brillouin
scattering an optical field generates a co-propagating acoustic field that in turn,
 changes the refractive index leading to Bragg scattering and thus a strongly reflected Stokes field. Conservation of energy requires
that the Stokes field is downshifted by the acoustic frequency which in fibres is typically around $11\, {\rm GHz}$. This reflected field can in 
turn be strong enough to create a second Stokes field leading to a cascade of multiple Stokes waves each an {\it identical} frequency separation.  
Cascaded Brillouin scattering is especially strong in fibre resonators where anywhere up to nearly 800 distinct 
Brillouin lines can be seen \cite{BumkiMin:dt} and more recently some groups have demonstrated that these lines can be phase-locked resulting in 
a 11Ghz pulse train (see \cite{Loranger:2012jn} and \cite{Buttner:2014cv}). Given the importance of and growing interest in Brillouin resonators it is 
perhaps surprising that a detailed theoretical analysis of cascaded Brillouin scattering has not been performed.  We present here a new analysis of 
the acoustic fields present in the fibre and show that this results in a new set of coupled equations for the amplitudes of the optical fields. Moreover 
this analysis shows that previous researchers have incorrectly generalised the equations for Brillouin scattering when trying to analyse cascaded Brillouin generation
leading to flawed results. We start by summarising the well known physics of simulated Brillouin scattering before extending it to cascaded Brillouin
scattering and finally discuss the differences between our work and previous analyses.

\section{Standard Analysis of Brillouin Scattering}
SBS is a three wave process in which a strong pump field $P$ creates a co-propagating phonon field $Q$ and a backwards propagating
Stokes field $B_0$ that is down shifted by the acoustic resonance frequency $\Omega$ of $\sim 11 {\rm GHz}$. For the case of a single Stokes
wave the density fluctuations $\rho$ in the fibre can be written as \cite{agrawal}
\begin{equation}
	\rho({\bf r},t) = F_A(x,y) Q(z,t) e^{-i (\Omega t - k_A z)} + c.c.
\label{acoustic}
\end{equation}
where $Q$ is the slowly varying acoustic field at frequency $\Omega$ propagating in the $+z$ direction with wave vector $k_A$. The acoustic
mode profile is given by $F_A$ and is assumed constant down the fibre. The phonon field $Q$ acts to couple the pump and Stoke's wave
leading to the standard coupled amplitude equations:
\begin{align}
\label{pump1}	\frac{\partial P}{\partial z} +\frac{1}{v_g} \frac{\partial P}{\partial t} & = -\alpha P + i \kappa B_0 Q \\
\label{stokes1}	-\frac{\partial B_0}{\partial z} +\frac{1}{v_g} \frac{\partial B_0}{\partial t} & = -\alpha B_0 + i \kappa P Q^* \\
\label{phonon1}	\frac{\partial Q}{\partial t} +{v_a} \frac{\partial Q}{\partial z} & = -\Gamma_A Q + \frac{i \delta}{A_{eff}} P B_0^*
\end{align}
where $v_g$ is the group velocity of the electric field at frequency $\omega_0$, $\alpha$ is the loss and $\kappa$ represents the coupling strength. The
acoustic wave propagates at a velocity of $v_a$ and is strongly damped by $\Gamma_A$ and is driven by the product of the electric fields through the
electrostriction term represented by  $\delta$. $A_{eff}$ represents the effective area of the fibre.

In analysing Brillouin generation it is usual to make the assumption that the fields are time-independent and furthermore the spatial derivative of $Q$ can be 
dropped due to the slow speed of the phonon field in which case Eq.~\eqref{phonon1} can
be solved for $Q$ and the usual equations for the power of the pump and stokes waves can be derived \cite{agrawal}
\begin{align}
	\label{pwer_pump} \frac{d I_p}{d z} & = -g_B I_p I_s - 2 \alpha I_p \\
	\label{pwer_stokes} -\frac{d I_s}{d z} & = g_B I_p I_s - 2 \alpha I_s
\end{align}
where $g_B$ is the standard Brillouin gain and $I_p = P P^*$ and $I_s= B_0 B_0^*$.  

In fibres and in particular resonant Brillouin lasers it is common that the Stokes wave becomes sufficiently intense that it creates a 
second phonon wave and thus
a second Stokes wave that is co-propagating with the pump and offset by twice the acoustic resonance frequency ($2 \Omega$). Denoting the second acoustic
wave by $R$ and the second Stokes wave as $F_0$ we get the coupled equations:
\begin{align}
\label{pump1.2}		\frac{\partial P}{\partial z} +\frac{1}{v_g} \frac{\partial P}{\partial t} & = -\alpha P + i \kappa B_0 Q \\
\label{stokes1.2}	-\frac{\partial B_0}{\partial z} +\frac{1}{v_g} \frac{\partial B_0}{\partial t} & = -\alpha B_0 + i \kappa  \left( P Q^*+ F_0 R \right) \\
\label{stokes2.2}	\frac{\partial F_0}{\partial z} +\frac{1}{v_g} \frac{\partial F_0}{\partial t} & = -\alpha F_0 + i \kappa  \left( B_0 R^* \right) \\
\label{phonon1.2}	\frac{\partial Q}{\partial t} +{v_a} \frac{\partial Q}{\partial z} & = -\Gamma_A Q + \frac{i \delta}{A_{eff}} P B_0^* \\
\label{phonon2.2}   \frac{\partial R}{\partial t} -{v_a} \frac{\partial R}{\partial z} & = -\Gamma_A R + \frac{i \delta}{A_{eff}} B_0 F_0^*
\end{align}
Again in the time-independent case and when the spatial dervatives of the phonon waves can be neglected we can solve for $Q$ and $R$ as
\begin{align}
	Q &= i \frac{\delta}{\Gamma_A A_{eff}} P B_0^* \\
	R &= i \frac{\delta}{\Gamma_A A_{eff}} B_0 F_0^*.
\end{align}
Substituting the expressions for $Q$ and $R$ into the Eq.~\eqref{pump1.2}--\eqref{stokes2.2} and assuming continuous wave interactions leads to the coupled power 
equations
\begin{align}
	\label{power_pump} \frac{d I_p}{d z} & = -g_B I_p I_{s0} - 2 \alpha I_p \\
	\label{power_stokes} -\frac{d I_{s0}}{d z} & = g_B \left( I_p I_{s0} -I_{s0} I_{s1} \right) - 2 \alpha I_{s0} \\
	\label{power_stokes2} \frac{d I_{s1}}{d z} & = g_B \left(I_{s0} I_{s1} \right) - 2 \alpha I_{s1} 
\end{align}
where $I_p$ is the power in the pump beam, $I_{s0}$ is the power in the backwards propagating first Stokes wave and $I_{s1}$ is the power in 
the forward propagating 2nd Stokes beam.  We emphasise here that the correct equations are Eqs.~\eqref{pump1.2}--\eqref{phonon2.2} rather than the 
coupled power equations which do not describe the full dynamics of the system (see Ref.~\cite{Randoux:1995cc} for example). Furthermore once a third 
stokes wave is introduced, then as we will show below, it is no longer possible to write down a set of  coupled power equations and thus any analysis that 
begins with such a set of equations is wrong. 

\section{Generation of Multiple Brillouin Lines}
Continuing the analysis from the previous section suppose that there are now $2N -1$ Brillouin lines generated in a fibre ($N$ can be over 390 see for example
Ref. \cite{BumkiMin:dt} ). The fundamental question is how many acoustic waves are present in the fibre? Since each pair of lines both generate and are coupled 
by an acoustic wave it might initially appear that there are $2N-1$ acoustic waves present of which $N$ are propagating in the forward direction and $N-1$
are propagating in the backward direction. Conservation of energy and momentum implies that the frequency $\Omega_j$ of the $jth$ acoustic wave will be given by
Stokes shift between the $jth$ and $(j+1) th$ electromagnetic wave. Thus we can write the total acoustic perturbation in the fibre as
\begin{align}
\label{phsum}	\rho(z,t) = F_A(x,y) \left( \sum_j Q_j(z,t) e^{-i (\Omega_j t - k_A(\Omega_j) z)} + \sum_m R_m(z,t) e^{-i (\Omega_m t + k_A(\Omega_m) z)} \right) +c.c.
\end{align}
where $Q_j$ and $R_m$ represent the forward and backward propagating acoustic waves respectively. 

In a fibre or waveguide the Brillouin gain is relatively narrow ($\sim 17\,{\rm MHz}$ in silica for example) compared to the wavelength shift ($11\,{\rm GHz}$) 
and so the standard assumption is that all of the Brillouin lines are {\it equally spaced in frequency } implying that
\begin{align}
	\Omega_l = \Omega,  \label{crucial} \quad l=1,2,... ,2N-1 
\end{align}
and thus the sums in Eq.~\eqref{phsum} each collapse to a single term leading to the simplified expression
\begin{align}
	\rho(z,t) = F_A(x,y) \left( Q(z,t) e^{-i (\Omega t - k_A(\Omega) z)} + R(z,t) e^{-i (\Omega t + k_A(\Omega) z)} \right) +c.c.
\end{align}
for the acoustic field. We note that measurements of the Brillouin shift in fibres for multiple cascaded Brillouin lines has found that Eq.~\eqref{crucial} is
correct\cite{Liu:2012ek}.

Eq.~\eqref{crucial} represents the crux of our argument since if Eq.~\eqref{crucial} is correct then there are {\it only 2} acoustic waves in the fibre: one
propagating in the forward direction and one in the backwards direction. Looking in the literature we have found only one paper
by C. Montes \cite{Montes:1985wn} published in 1985 that discusses this situation (in the context of plasma physics) which he calls the strongly coupled case (see Eqs. 47--50
of Ref~\cite{Montes:1985wn}). More recently Ogusu \cite{Ogusu:2002vr} for example keeps multiple acoustic waves despite the fact that he assumes that 
each Brillouin shift is identical. This approach is followed by Buttner et al. \cite{Buttner:2014cv} who in their supplementary information (Eq. S1) state that the each Brillouin
wave is shifted by an identical amount. A third recent example is  Yuan et al. \cite{Yuan:2014et} who implicitly adopt this model in their numerical analysis of their results. 

We argue here that if there is only a single acoustic wave propagating in each direction in the fibre then it must be driven by multiple sets of interacting fields. 
This has important implications for the derivation of the correct set of equations to describe cascaded
Brillouin scattering. In particular it implies that unlike the case for 2 or 3 interacting waves coupled power equations cannot be
used and instead coupled equations for the fields are required and the correct equations which are a generalisation of the work of Montes \cite{Montes:1985wn}
are presented below.

In line with the notation introduced above we denote the pump beam by $P$, the backwards 
propagating Stokes waves by $B_i$ and the forward propagating Stokes waves by $F_i$ where $i$ runs from 0 to $N$. Using this notation the 
correct equations are:
\begin{align}
	\label{p0}	\frac{\partial P}{\partial z} +\frac{1}{v_g} \frac{\partial P}{\partial t} & = -\alpha P + i \kappa B_0 Q \\
	\label{s0}	-\frac{\partial B_0}{\partial z} -\frac{1}{v_g} \frac{\partial B_0}{\partial t} & = -\alpha B_0 + i \kappa  \left( P Q^*+ F_0 R \right) \\
	\vdots  \nonumber \\
	\label{bi}	-\frac{\partial B_i}{\partial z} +\frac{1}{v_g} \frac{\partial B_i}{\partial t} & = -\alpha B_i + i \kappa  \left( F_{i-1} Q^* +F_i R \right) \\
	\label{fi}	\frac{\partial F_i}{\partial z} +\frac{1}{v_g} \frac{\partial F_i}{\partial t} & = -\alpha F_i + i \kappa  \left( B_i R^* +B_{i+1} Q \right) \\
	\vdots  \nonumber \\
	\label{fn}	\frac{\partial F_N}{\partial z} +\frac{1}{v_g} \frac{\partial F_N}{\partial t} & = -\alpha F_N + i \kappa  \left( B_i R^*  \right) 
\end{align}
for the optical fields (assuming an even number of fields). While the acoustic waves satisfy 
\begin{align}
\label{ph1}	\frac{\partial Q}{\partial t} +{v_a} \frac{\partial Q}{\partial z} & = -\Gamma_A Q + \frac{i \delta}{A_{eff}} 
	\left(P B_0^* +\sum_{j=1}^{N-1} F_j B_{j+1}^* \right)\\
\label{ph2}   \frac{\partial R}{\partial t} -{v_a} \frac{\partial R}{\partial z} & = -\Gamma_A R + \frac{i \delta}{A_{eff}} \left( \sum_{j=0}^N B_j F_j^* \right)
\end{align}
Making the standard approximations for the acoustic waves leads to the equations for $Q$ and $R$:
\begin{align}
	\label{q1} Q & = i \frac{\delta}{\Gamma_A A_{eff}} \left( P B_0^* + \sum_{j=0}^{N-1} F_j B_{j+1}^* \right) \\
	\label{qr1} R & = i \frac{\delta}{\Gamma_A A_{eff}} \left(  \sum_{j=0}^{N} B_j F_{j}^* \right) 
\end{align}
which can then be substituted back into the equations for the fields [Eq.~\eqref{p0}--Eq.~\eqref{fn}] if desired. In that case it can be easily shown that
the coupled equations conserve the total photon flux through the system.

\section{Discussion and Conclusions}
The new equations Eq.~\eqref{p0}--\eqref{ph2} are the correct and obvious generalisation of the coupled equations for the case of two 
Stokes waves when there are only two acoustic waves propagating in the fibre. However in the literature most authors have chosen, incorrectly, 
to generalised  the coupled power equations [Eq.~\eqref{power_pump}
--Eq~\eqref{power_stokes2}] or equivalently to include a separate acoustic wave for each new pair of Stokes waves. Not surprisingly, since
this fundamental physical picture of the interaction is wrong, it leads to an incorrect set of equations and thus the resulting analysis 
cannot be trusted. We emphasise here the point made by Montes in his original article that in case of more than 3 interacting optical
fields it is not possible to write down coupled equations for the evolution of the power in the fields and rather the coupled amplitude
equations must be used. 

A direct consequence of the presence of only two acoustic fields is that all optical fields are coupled together via the two phonon fields. 
An effect of this is to alter the threshold required to see multiple lines since each field has a growth term proportional to
the pump $P$ (forward propagating fields) or its complex conjugate (backward propagating fields). Thus one would expect that there 
would be more Brillouin lines in a resonator than would be expected using the analysis of Ogusu for example. We are currently looking at 
numerically solving the full coupled equations and will present a more detailed numerical analysis of the difference between the two sets of
equations in a later paper.

We also note that it is relatively trivial to extend the above equations to include Kerr nonlinear terms and also to describe ring resonators or Fabry-Perot cavities
such as those treated by both Buttner {\it et al.} and by Ogusu. Treating a Fabry-Perot cavity 
requires that the number of optical fields is  doubled since each Stoke wave is reflected by the cavity mirrors giving rise to a forward and backward propagating wave 
at the same frequency. However even in this case the number of acoustic waves remains fixed at 2 and so we do not treat this case any further. 

In conclusion we have highlighted a common error in the analysis of cascaded Brillouin generation in optical fibres and presented the correct 
analysis based on correctly identifying the acoustic fields in the fibre (following earlier analysis by Montes). We hope that this work will be of use to 
other researchers working in the field.

\section*{Funding Information}
Marsden fund of the royal society of New Zealand.

\bibliographystyle{IEEEtran}

\end{document}